\documentclass[a4paper,12pt]{article}
\DeclareMathSizes{12}{12.5}{10}{10}
\usepackage[left=2.3cm,bottom=3cm,right=2.3cm,top=3cm]{geometry}
\usepackage{overpic,youngtab}
\usepackage{subfigure}
\usepackage[latin,english]{babel}
\usepackage{amsmath}
\usepackage{verbatim}
\usepackage{amssymb}
\usepackage{footmisc}
\usepackage{latexsym}
\usepackage{epsfig}
\usepackage{epstopdf}
\usepackage{graphics,psfrag,rotating}
\usepackage{graphicx}
\usepackage{dcolumn}
\usepackage{float}
\usepackage{pdflscape}
\usepackage{array}
\usepackage{booktabs}
\usepackage{amscd}
\usepackage{fancybox}
\usepackage{color}
\usepackage[colorlinks=true,citecolor=blue,,linktocpage=true,linkcolor=blue,urlcolor=black]{hyperref}
\usepackage{mathabx}

\definecolor{ogreen}{rgb}{0,0.7,0}

\DeclareGraphicsExtensions{.pdf,.png,.jpg,.gif,.jpeg}
\def\be{\begin{equation}}

\def\ee{\end{equation}}
\def\bea#1\eea{\begin{align}#1\end{align}}
\def\pd{\partial}

\def\a{\alpha}
\def\b{\beta}

\def\d{\delta}

\def\k{\kappa}
\def\m{\mu}
\def\n{\nu}

\def\l{\lambda}

\def\r{\rho}

\def\s{\sigma}

\def\bi{\begin{itemize}}
	\def\ei{\end{itemize}}

\usepackage{authblk}
\usepackage{setspace}

\def\bea#1\eea{\begin{align}#1\end{align}}
\def\pd{\partial}
\def\a{\alpha}

\def\b{\beta}

\def\d{\delta}

\def\k{\kappa}
\def\m{\mu}
\def\n{\nu}

\def\l{\lambda}

\def\r{\rho}

\def\s{\sigma}

\def\bi{\begin{itemize}}
	\def\ei{\end{itemize}}

\def\N{\nabla}

\usepackage{authblk}
\usepackage{setspace}

\usepackage{tikz}
\makeatletter
\newcommand\mathcircled[1]{%
	\mathpalette\@mathcircled{#1}%
}
\newcommand\@mathcircled[2]{%
	\tikz[baseline=(math.base)] \node[draw,circle,inner sep=1pt] (math) {$\m@th#1#2$};%
}
\makeatother

\begin{document}

	%%%%%%%%%%%%%%%%%%%%%%%%%%%%%%%%%%%%%%%%%%%%%%%%%%%%%%%%%%%%%%%%%%%%%%%%%%%%%

	\vspace*{-1cm}
	{\flushleft
		{{FTUAM-18-16}}
		\hfill{{IFT-UAM/CSIC-18-65}}}
	\vskip 1.5cm
	\begin{center}
		{\LARGE\bf Massive Unimodular Gravity}\\[3mm]
		\vskip .3cm
		
	\end{center}
	\vskip 0.5  cm
	\begin{center}
		{\Large Enrique Alvarez, Jesus Anero, Guillermo Milans del Bosch and Raquel Santos-Garcia }$^{~a}$		\\
		\vskip .7cm
		{
			$^{a}$Departamento de F\'isica Te\'orica and Instituto de F\'{\i}sica Te\'orica (IFT-UAM/CSIC),\\
			Universidad Aut\'onoma de Madrid, Cantoblanco, 28049, Madrid, Spain\\
			\vskip .1cm

			\vskip .5cm
			\begin{minipage}[l]{.9\textwidth}
				\begin{center} 
					\textit{E-mail:} 
					\tt{enrique.alvarez@uam.es},
					\tt{jesusanero@gmail.com},
					\tt{guillermo.milans@csic.es},
					\tt{raquel.santosg@uam.es}
				\end{center}
			\end{minipage}
		}
	\end{center}
	\thispagestyle{empty}
	
	\begin{abstract}
		\noindent
		A ghost free massive deformation of unimodular gravity (UG), in the spirit of {\em mimetic massive gravity}, is shown to exist. This construction avoids the no-go theorem for a Fierz-Pauli type of mass term in UG by giving up on Lorentz invariance. In our framework, the mimetic degree of freedom vanishes on-shell. 
	\end{abstract}
	%\maketitle
	\newpage
	\tableofcontents
	\thispagestyle{empty}

	\newpage
		\setcounter{page}{1}
%%%%%%%%%%%%%%%%%%%%%%%%%%%%%%%%%%%%%%%%%%%%%%%%%%%%%%%%%%%%%%%%%%%%%%%%%%%%%%%%%%%%%%%%%%%%
	\section{Introduction}
%%%%%%%%%%%%%%%%%%%%%%%%%%%%%%%%%%%%%%%%%%%%%%%%%%%%%%%%%%%%%%%%%%%%%%%%%%%%%%%%%%%%%
Unimodular gravity (UG) \cite{Unruh:1988in} is a variant of General relativity (GR) where the metric is assumed to be unimodular, that is, to have unit determinant ({\em confer} \cite{AlvarezBGV2} for a brief expos\'e). The main interest of UG lies in the way vacuum energy affects gravitation. Namely, vacuum energy does not weigh at all. This solves one half of the cosmological constant problem (see for example \cite{Alvarez} and references therein), namely, why it is not much bigger than observed. Incidentally, this aspect of the problem was first pointed out by Pauli, which proposed that vacuum energy should be ignored when speaking of gravitation \cite{Pauli}. A crucial result is that given the classical equivalence of the equations of motion of both theories, the number of degrees of freedom propagated by UG matches the ones of GR,  namely, a single massless graviton \cite{vanderBij:1981ym, Herrero-Valea:2018ilg}. This naturally raises the question of whether UG is fully equivalent to GR (apart from the aforementioned role of the cosmological constant) or if there is some physical phenomenon that can distinguish both theories (see \cite{ASGM,percacci},  and references therein).
\par
One of the distinctive features of UG that interests us for this paper is that there is no massive deformation of it in a flat background \cite{AlvarezBGV}. This no-go theorem assumes, besides Lorentz invariance, unitarity. That is, any possible Lorentz invariant mass term in UG implies a ghost. These ideas have also been analyzed in \cite{Bonifacio:2015rea}. This is very much unlike the linear limit of GR, which admits a massive deformation, which is besides uniquely determined by Lorentz invariance and unitarity: the Fierz-Pauli mass term \cite{FierzPauli}. 
\par 
In this paper, we shall present a way out of the above no-go theorem, albeit one that violates Lorentz invariance, but which we believe to be interesting nonetheless as Lorentz breaking massive deformations are the only possible ones in UG.  The present model is a small variation of the mimetic one of Chamseddine and Mukhanov \cite{Chamseddine,Chamseddine2,Chamseddine3}, but as a result of the unimodularity constraint, the mimetic degree of freedom is lost on-shell. This tackles the main difference between UG and GR since the mimetic degree of freedom is related to the trace of the equations of motion. In this sense, the unimodular counterpart of mimetic gravity provides an explicit example where the counting of the dynamical degrees of freedom differs from the usual mimetic gravity embedded in GR. 
\par
In \cite{Chamseddine} mimetic matter is introduced, originally as a candidate for dark matter, by considering two conformally related metrics in spacetime
\be
\hat{g}_{\m\n}\equiv \Omega^2(x) g_{\m\n},
\ee
where the conformal factor is determined by a kinetic energy of a scalar field
\be
\Lambda^4 \Omega^2\equiv g^{\m\n}\pd_\m\phi\pd_\n\phi.
\label{def}
\ee
Here, $\Lambda$ is just an energy scale. Using \eqref{def}, it follows that the scalar field satisfies the following relation
\be \hat{g}^{\m\n}\partial_\m\phi\partial_\n\phi=\frac{1}{\Omega^2}g^{\m\n}\partial_\m\phi\partial_\n\phi=\Lambda^4,\ee
that is, its kinetic energy is constrained to be a constant. 
The key point that we want to exploit is that it is possible to impose the extra constraint\footnote{In a \textit{Diff} invariant theory, this equality would not make sense since we would be equating a spacetime scalar in the left hand side with a scalar density in the right hand side. However, when dealing with \textit{TDiff} invariant theories as in our case both sides are \textit{TDiff} scalars because the determinant is inert under transverse diffeomorphisms.\label{footnote1}}
\be
g^{\a\b}\pd_\a\phi \pd_\b \phi= \Lambda^4 g^{- \frac{1}{4}},
\label{unimdef}
\ee
with $g\equiv \text{det}\,g_{\m\n}$, in such a way that the physical metric is unimodular
\be
\hat{g}_{\m\n}\equiv g^{-\frac{1}{4}} g_{\m\n} \hspace{.2cm}\Longrightarrow \hspace{.2cm}\hat{g}\equiv \text{det}\,\hat{g}_{\m\n}=1. \label{unimodular metric}
\ee
Therefore, the mimetic construction is compatible with the unimodular constraint. First of all, let us notice that this condition is Weyl invariant in the sense that it does not change under
\be
g_{\m\n}\rightarrow \omega^2(x) g_{\m\n},
\label{weylconstr}
\ee
Second, it is straightforward to see that \eqref{unimdef} implies that the physical norm of the kinetic energy of the scalar field still equals $\Lambda^4$.
%\be
%\hat{g}^{\m\n} \pd_\m\phi\pd_\n\phi= g^{\frac 1 4 }\pd_\m\phi\pd_\n\phi =  \Lambda^4. \label{UP}
%\ee
After this change, the symmetry group of the theory consists of the Weyl symmetry that we get by construction \eqref{weylconstr} together with the transverse diffeomorphism invariance in order to preserve the unimodularity condition, that is, {\em WTDiff}\footnote{Let us note that this description of unimodular metrics is equivalent to Henneaux and Teitelboim's approach \cite{Henneaux:1989zc} via a lagrange multiplier. This is reanalyzed in \cite{Alvarez:2005iy} where  it can be seen that in the equation of motion for the lagrange multiplier, the volume form is supposed to be equal to an exact differential. This is impossible in general, because the volume form generates the top cohomology class, that is, it is closed, but not exact. The only way out is to take $g=1$, in which case the equation has the Poincare solution as in $\mathbb{R}^n$. But then, when fixing $g=1$, the symmetry group is reduced to the volume preserving (that is, transverse) diffreomorphisms. In this sense, our approach, although non-covariant from the beginning, leads to the same symmetries at the level of the equations of motion.} \cite{AlvarezBGV}.
\par
The aim of this paper is to apply the massive gravity model introduced in \cite{Chamseddine2,Chamseddine3} with the further constraint of UG \eqref{unimodular metric}, to construct an unimodular massive gravity. The second section of the paper is devoted to such construction. We also discuss the position of this model in the general setup of \cite{Dubovsky}. 
%%%%%%%%%%%%%%%%%%%%%%%%%%%%%%%%%%%%%%%%%%%%%%%%%%%%%%%%%%%%%%%%%%%%%%%%%%%%%%%%%%%%%%%%%%%%
\section{Unimodular massive gravity}
%%%%%%%%%%%%%%%%%%%%%%%%%%%%%%%%%%%%%%%%%%%%%%%%%%%%%%%%%%%%%%%%%%%%%%%%%%%%%%%%%%%%%%%%%%%%

In \cite{Chamseddine2,Chamseddine3} a new version of massive gravity is proposed. It is further argued that it is free of the Boulware-Deser (BD) \cite{Boulware} ghost to all orders of perturbation theory. The main idea is to introduce the Brout-Englert-Higgs mechanism for gravity through four scalar fields $\phi^a,\, a=0\ldots 3$ that acquire a vacuum expectation value. In flat space
\be
\bar{\phi}^a\equiv \langle \phi^a\rangle=\Lambda^2 \d^a_\m x^\m\equiv \Lambda^2x^a,
\ee
and the perturbations of these scalar fields (which will be Goldstone bosons \cite{Dubovsky}) are defined as
\be
\phi^a\rightarrow \bar{\phi}^a+\xi^a.
\ee
In \cite{Chamseddine2} there are four independent fields $\xi^a$ and latin indices are raised and lowered with the Minkowski metric $\eta_{ab}$. The three scalars $\xi^i$ , where $i=1\ldots 3$, give mass to the graviton, while the fourth one, $\xi^0$, which in other models usually becomes a ghost, is now constrained to have unit norm because of the mimetic constraint which is consistent with the unimodular projection
\be\label{unit}
\hat{g}^{\m\n} \pd_\m\phi^0\pd_\n\phi^0=\Lambda^4.
\ee

In our case the symmetry of UG is not the full diffeomorphism invariance, but only those {\em volume preserving} diffeomorphisms \cite{Arnold}, whose component connected with the identity is generated by {\em transverse} vector fields; that is, those that obey
\be
\pd_\m \xi^\m=0.
\ee
We have coined the name {\em TDiff} for those. Besides, we have a Weyl invariance appearing by construction of the conformal factor, so that the symmetry group of the theory is now {\em WTDiff} \cite{AlvarezBGV}. 
\par
Then, the induced {\em metric perturbations}\footnote{Our conventions for the flat metric are $\eta^{ab}=\mbox{diag}(1,-1,-1,-1)$}
\be
\kappa \hat{h}^{ab}\equiv \Lambda^{-4} \hat{g}^{\m\n}\pd_\m\phi^a\pd_\n\phi^b-\eta^{ab}\label{h},
\ee    
which in our case are transverse diffeomorphism scalars obeying
\be
\hat{h}^{00}=0
\ee    
exactly, as long as the constraint \eqref{unit} is maintained. 
\par
In \cite{Chamseddine2, Chamseddine3} the action for massive gravity that is postulated reads
\be
S= \int \sqrt{|\hat{g}|}\,d^4 x\,\bigg\{-{1\over 2\kappa^2} R[\hat{g}]+{m^2\over 8}\left({\hat{h}^2\over 2}-\hat{h}^{ab}\hat{h}_{ab}\right)+\l\left(\hat{g}^{\m\n} \pd_\m\phi^0 \pd_\n\phi^0-\Lambda^4\right)\bigg\}
\ee
where $\lambda$, which is dimensionless and of first order in the perturbation scheme, is denoted as {\em mimetic matter} and presented as a candidate for dark matter. 

The remarkable fact is that this mass term is not of the Fierz-Pauli type, which is the only one that is unitary in a flat background. As we shall see, this is due to the fact that the extra condition imposed through a Lagrange multiplier breaks Lorentz invariance. 
\par
Now, taking into account the unimodular transformation \eqref{unimodular metric}, the action for the {\em WTDiff} version of this theory reads
\bea
S_{\text{\tiny{\em WTDiff}}}&= \int d^4 x\,\bigg\{-{1\over 2\kappa^2} |g|^{\frac{1}{4}}\left(R+\frac{3}{32}\frac{\nabla_\m g\nabla^\m g}{g^2}\right)+\l\left(|g|^{\frac{1}{4}}g^{\m\n} \pd_\m\phi^0 \pd_\n\phi^0-\Lambda^4\right)\nonumber\\
&+{m^2\over 8}\left(\frac{\Lambda^{-8}}{\k^2}|g|^{\frac{1}{2}}g^{\m\n}g^{\a\b}\left(\frac{1}{2}\pd_\m\phi^a\pd_\n\phi_a\pd_\a\phi^b\pd_\b\phi_b-\pd_\m\phi^a\pd_\a\phi_a\pd_\n\phi^b\pd_\b\phi_b\right)\right.\nonumber\\
&\left.-\frac{2\Lambda^{-4}}{\k^2}|g|^{\frac{1}{4}}g^{\m\n}\pd_\m\phi^a\pd_\n\phi_a+\frac{4}{\k^2}\right)\bigg\}.
\eea
We define the covariant derivative acting on the metric determinant as ordinary partial differentiation. Let us note that in this case the metric determinant behaves as an ordinary scalar instead of a scalar density. As mentioned in footnote \footref{footnote1} this is due to the fact that we are dealing with a theory that is \textit{TDiff} invariant instead of \textit{Diff} invariant and hence the determinant is always inert under transverse diffeomorphisms (the variation of the metric determinant is controled by the longitudinal part of the usual diffeomorphisms). 
\par
The equation of motion (eom) for the metric takes the form
\bea
& \left(-{1\over 4}\,R g^{\r\s}+R^{\r\s}\right)-\dfrac{7}{32}\left(\dfrac{\nabla^\r g \nabla^\s g}{g^2} -\dfrac 1 4 \dfrac{(\nabla g)^2}{g^2} g^{\r\s} \right) + \dfrac 1 4 \left( \dfrac{\nabla^\r  \nabla^\s g}{g} -\dfrac 1 4 \dfrac{\Box g}{g} g^{\r\s}\right)
\nonumber\\
&-2\kappa^2 \left(g^{\m\r}g^{\n\s}-\frac{1}{4}g^{\m\n}g^{\r\s}\right)\left\{\frac{m^2}{4\k^2}\bigg[\Lambda^{-8}|g|^{\frac{1}{4}}g^{\a\b}\left(\frac{1}{2}\pd_\m\phi^a\pd_\n\phi_a\pd_\a\phi^b\pd_\b\phi_b-\pd_\m\phi^a\pd_\a\phi_a\pd_\n\phi^b\pd_\b\phi_b\right)\right.\nonumber\\
&\left.-\Lambda^{-4}\pd_\m\phi^a\pd_\n\phi_a\bigg]+\l\pd_\m\phi^0\pd_\n\phi^0\right\}=0.
\eea
Let us make some remarks at this point. Due to Weyl invariance, these equations are traceless. Besides, the equations coincide with the traceless part of the gravitational equation in \cite{Chamseddine2,Chamseddine3} in the gauge $g=1$. In that case, it is known that due to the Bianchi identities we can recover the trace of the gravitational eom. UG is then equivalent to GR with a cosmological constant given by the integration constant coming from the Bianchi identities. In the gauge $g\neq 1$, however, the equations are not related to the ones in \cite{Chamseddine2,Chamseddine3}, as they are the Weyl transformed version of Einstein's equations with a cosmological constant as shown in appendix \ref{A} for purely gravitational theories. Nevertheless, in the absence of terms breaking these symmetries, we can always gauge fix the extra Weyl symmetry so that the classical eom of the theories are equivalent. 
\par
The eom for the scalar fields reads
\bea
&\pd_\m\bigg\{2\eta^{a 0} |g|^{\frac{1}{4}} g^{\m\n}\l\pd_\n\phi^0-\frac{m^2\Lambda^{-4}}{2\k^2}|g|^{\frac{1}{4}} g^{\m\n}\pd_\n\phi^a\nonumber\\
&+\frac{m^2}{8}\left[2g^{\m\n}g^{\a\b}|g|^{\frac{1}{2}}\frac{\Lambda^{-8}}{\k^2}\left(\pd_\n\phi^a\pd_\a\phi_b\pd_\b\phi^b-\pd_\a\phi^a\pd_\n\phi_b\pd_\b\phi^b-\pd_\a\phi^b\pd_\n\phi_b\pd_\b\phi^a\right)\right]\bigg\}=0.\eea
Obviously, the eom for $\l$ reproduces the constraint \eqref{unit}, so that
\be \hat{h}^{00}=0.\ee
%
%These equations enjoy Weyl invariance besides {\em TDiff}. This is what we have dubbed {\em WTDiff} in \cite{AlvarezBGV}. It is  almost always convenient to fix the Weyl gauge symmetry setting
%\be
%g=1,
%\ee
%which in the linearized theory translates into 
%\be
%h\equiv h^\m_\m=0.
%\ee
%In this gauge, the eom for the metric reads
%\bea
%&\left(g^{\m\r}g^{\n\s}-\frac{1}{4}g^{\m\n}g^{\r\s}\right)\left\{\frac{m^2}{8}\bigg[\frac{2\Lambda^{-8}}{\k^2}g^{\a\b}\left(\frac{1}{2}\pd_\m\phi^a\pd_\n\phi_a\pd_\a\phi^b\pd_\b\phi_b-\pd_\m\phi^a\pd_\a\phi_a\pd_\n\phi^b\pd_\b\phi_b\right)\right.\nonumber\\
%&\left.-2\frac{\Lambda^{-4}}{\k^2}\pd_\m\phi^a\pd_\n\phi_a\bigg]+\l\pd_\m\phi^0\pd_\n\phi_0\right\}+{1\over 2\kappa^2} \left({1\over 4}\,R g^{\r\s}-R^{\r\s}\right)=0.
%\eea
%Note that this equation is traceless, as expected due to {\em TDiff} invariance. For the scalar fields we obtain
%\bea
%&\pd_\m\bigg\{2\eta^{a 0}g^{\m\n}\l\pd_\n\phi^0-\frac{m^2\Lambda^{-4}}{2\k^2}g^{\m\n}\pd_\n\phi^a+\nonumber\\
%&+\frac{m^2}{8}\left[2g^{\m\n}g^{\a\b}\frac{\Lambda^{-8}}{\k^2}\left(\pd_\n\phi^a\pd_\a\phi_b\pd_\b\phi^b-\pd_\a\phi^a\pd_\n\phi_b\pd_\b\phi^b-\pd_\a\phi^b\pd_\n\phi_b\pd_\b\phi^a\right)\right]\bigg\}=0.\eea
Let us now in fact linearize those equations by writing
\bea
&g_{\m\n}=\eta_{\m\n}+\kappa h_{\m\n}\nonumber\\
&\phi^a= \Lambda^2 x^a+\xi^a
\eea
with the {\em TDiff} constraint
\be 
\pd_a \xi^a=0.
\ee
It is plain that now, greek indices are raised and lowered with the Minkowski metric $\eta_{\m\n}$.
The linearized gravitational eom then reads
\bea
&{\kappa \over 2}\left[\pd^\l\pd^\m h_{\l}^\n+\pd^\l\pd^\n h_{\l}^\m-\Box h^{\m\n}-\frac{1}{2}\pd^\l\pd^\s h_{\l\s}\eta^{\m\n} + \dfrac 3 8 \eta^{\mu \n} \Box h -\dfrac 1 2 \partial^\m \partial^\n h \right]+ \frac{m^2}{8} \eta^{\m \n}\left(\kappa h- 2 \Lambda^{-2} \pd^\a\xi_\a\right)\nonumber\\
&-\frac{m^2}{2}\left(\kappa h^{\m\n}- \Lambda^{-2}(\pd^\m\xi^\n+\pd^\n\xi^\m)\right)-2 \l \kappa^2\Lambda^4\left(\eta^{\m 0}\eta^{\n 0}- \frac{1}{4}\eta^{\m \n}\right)=0
\label{graveom}
\eea
where the relation between greek and latin indices comes from $ x^a \equiv \d_\m^a x^\m$.
It should be noticed that the second member of this equation, proportional to the mass parameter, breaks the {\em TDiff} gauge symmetry. Tracelessness of the eom is still attained as $\pd_\m \xi^\m = 0$. Now we can linearize \eqref{h} 
\bea
\kappa\hat{h}^{ab}&=\Lambda^{-4}\left(\eta^{\m\n}-\kappa h^{\m\n} + \dfrac 1 4 \kappa \, h \, \eta^{\m\n} \right)\left(\Lambda^2\d^a_\m +\pd_\m\xi^a\right)\left(\Lambda^2\d^b_\n+\pd_\n\xi^\b\right)-\eta^{ab}= \nonumber \\
&=-\kappa h^{ab}+  \dfrac 1 4 \kappa \, h \, \eta^{ab} +\Lambda^{-2}\left(\pd^a\xi^b+\pd^b\xi^a\right)
\eea
Nevertheless, we can reabsorb the terms containing $\xi^\m$ by a field redefinition
\be
h^{\m\n}-{1\over \kappa\Lambda^2}\left(\pd^\m\xi^\n+\pd^\n\xi^\m\right)\rightarrow  h^{\m\n}
\label{redefinition}
\ee
leading to
\bea
&\left[\pd^\l\pd^\m h_{\l}^\n+\pd^\l\pd^\n h_{\l}^\m-\Box h^{\m\n}-\frac{1}{2}\pd^\l\pd^\s h_{\l\s}\eta^{\m\n}+ \dfrac 3 8 \eta^{\mu \n} \Box h -\dfrac 1 2 \partial^\m \partial^\n h\right]-m^2 h^{\m\n} + \dfrac{m^2}{4}h \,  \eta^{\m\n} 
\nonumber\\
& -4\l\k\Lambda^4\left(\eta^{\m 0}\eta^{\n 0}  -\frac{1}{4}\eta^{\m \n}\right)=0\label{grav_eom}. \eea
The eom for the scalar field then yields (with $\pd_\m \xi^\m = 0$)
\be
\pd_\m\left[m^2 h^{\m \n}\d^a_\n -\dfrac{m^2}{4} \eta^{\m a} \, h +4\l\k\Lambda^4\eta^{a 0}\eta^{\m 0}\right]=0\label{scalar_eom}.\ee

The eom for the lagrange multiplier enforces that $\hat{h}^{00}=0$ exactly\footnote{We thank the anonymous referee for pointing out the trace piece in the constraint.}, so that to linear order and redefining the field again as in \eqref{redefinition} we get
\be
-\kappa h^{00}+\dfrac 1 4 \kappa h \, \eta^{00} = 0
\label{constraint}
\ee
Using the linearized Bianchi identity an integrability condition for the above equation has to be fullfilled
\be
\pd_\m\left[\dfrac 1 2 \pd^\l\pd^\s h_{\l\s}\eta^{\m\n} -\dfrac 1 8 \Box h \,  \eta^{\m\n} + \dfrac{m^2}{4} h \, \eta^{\m\n}-m^2 h^{\m \n}-4\l\k\Lambda^4\left(\eta^{\m 0}\eta^{\n 0}- \frac{1}{4}\eta^{\m \n}\right)\right]=0\label{Bianchi}.\ee
Combining \eqref{scalar_eom} with \eqref{Bianchi} leads to
\be
\pd_\m\left[\dfrac 1 2 \pd^\l\pd^\s h_{\l\s}\eta^{\m\n} -\dfrac 1 8 \Box h \,  \eta^{\m\n} +\l\k\Lambda^4\eta^{\m \n}\right]=0
\ee
so that we find the relation
\be
\pd^\l\pd^\s h_{\l\s} -\dfrac 1 4 \Box h +2 \l\k\Lambda^4 = C,
\label{Bianchiresult}
\ee
where $C$ is the integration constant arising when integrating the Bianchi identity. 
Deriving again the eom for the scalar field \eqref{scalar_eom} we get
\be
m^2 \pd_\m \pd_\n h^{\m\n} - \dfrac{m^2}{4} \Box h = - 4 \kappa \Lambda^4 \ddot{\l}.
\label{devscalar_eom}
\ee
Putting together \eqref{Bianchiresult} and \eqref{devscalar_eom} we arrive to
\be
\ddot{\l} - \dfrac{m^2}{2} \l +  \dfrac{m^2}{4 \kappa \Lambda^4} C = 0.
\label{eomC}
\ee
In principle this equation has the dangerous sign in the mass term so that its solution will generically contain a runaway. Nevertheless, we can get another equation for $\l$ so that the solution is not in conflict with \cite{Chamseddine2,Chamseddine3}.
If we derive the scalar eom with respect to a different index one gets
\be
m^2 \pd_\m \pd_\l h^{\l \n} - \dfrac{m^2}{4} \pd_\m \pd^\n h + 4 \, \kappa \Lambda^4 \pd_\m \dot{\l} \eta^{\n0} = 0.
\label{dev2scalar_eom}
\ee
Introducing \eqref{devscalar_eom} and \eqref{dev2scalar_eom} in the gravitational equation of motion \eqref{graveom} we get
\bea
&- \dfrac{4 \kappa \Lambda^4}{m^2}\pd^\m \dot{\l} \eta^{\n0} - \dfrac{4 \kappa \Lambda^4}{m^2}\pd^\n \dot{\l} \eta^{\m0} - \Box h^{\mu\nu}  + \dfrac 1 4  \Box h \, \eta^{\m\n} +  \dfrac{2 \kappa \Lambda^4}{m^2} \ddot{\l} \, \eta^{\m\n} - m^2 h^{\m \n} + \dfrac{m^2}{4} h \, \eta^{\m\n} \nonumber \\
&-4\l\k\Lambda^4\left(\eta^{\m 0}\eta^{\n 0}  -\frac{1}{4}\eta^{\m \n}\right)=0.
\label{graveom2}
\eea
Taking into account the constraint $h_{00} = \dfrac 1 4 h $, the $00$ component of this equation yields
\be
\ddot{\l} + \dfrac{m^2}{2}\l = 0.
\label{Boxh}
\ee
Let us note, that this is precisely the eom for $\l$ found in \cite{Chamseddine2}, except for a factor in the mass term. The difference in the equations comes from the fact that given the extra Weyl invariance present in our model, the gravitational equations of motion \eqref{graveom2} correspond to the traceless part of the gravitational equations in the standard one. Nevertheless, in this case, when combined with \eqref{eomC} we get
\be
\l = \dfrac{C}{4 \kappa \Lambda^4}.
\label{C}
\ee
Moreover, combining this equation with \eqref{Boxh}, the constant of integration is fixed to zero. 
Instead of $\l$ being a dynamical degree of freedom, for {\em WTDiff} theories, the mimetic degree of freedom vanishes on-shell. This is not a surprise because, in \cite{Chamseddine},  $\l$ is related to $(G-T)$, the traces of the Einstein tensor and the energy-momentum tensor, which in UG is fixed to a constant of integration reintroduced via the Bianchi identities (moreover, the constant is fixed to zero in this particular case.)
% \textcolor{red}{Este ecuacion solo cambia en el factor del termino  $(\Box + m^2) h$, antes era $\dfrac 1 4$. Aunque la constraint si que fuese Weyl invariant, esta ecuacion no lo es y no cambia ninguna de las conclusiones.} \textcolor{red}{He puesto el argumento de antes aqui.}
% It is important to note that this equation is not Weyl invariant. In the same way that in \cite{Chamseddine2} the mass terms break {\em Diff} invariance, in this case, the mass terms break {\em TDiff} and Weyl invariance. Therefore, as we will see, the eom of {\em WTDiff} and {\em TDiff} are no longer equivalent because we cannot gauge fix the extra Weyl invariance.  
The $ij$ component then yields
\be
-(\Box + m^2) h_{ij} + \dfrac 1 4 (\Box + m^2) h \eta_{ij}+2 \dfrac{ \kappa \Lambda^4}{m^2} \ddot{\l} \eta_{ij}+ \l\k\Lambda^4 \eta_{ij} = 0.
\ee
% \be
% - (\Box + m^2) h^{i}_i  + \dfrac 3 4 (\Box + m^2) h = 0 
% \ee
% We know 
With the constraint \eqref{constraint} $h^i_i ={3 \over 4 } h$ and using \eqref{Boxh} we can rewrite the previous equation as
% \be
% - (\Box + m^2) h^{i}_i  + \dfrac 3 4 (\Box + m^2) h = -(\Box + m^2) h
% \ee
%which is exactly the same as \eqref{graveom2}. Using both equations
\be
(\Box + m^2) h_{ij} - \dfrac 1 3 (\Box + m^2) h^k_k  \, \eta_{ij} =  (\Box + m^2) h^T_{ij} = 0,
\ee
which describes the 5 massive degrees of freedom.
The $0-i$ component finally reads
\bea
&\left(\Box +{ m^2}\right)\,h_{0i} = 0\label{h0i0}.\eea
Let us focus on these mixed $0-i$ components. Our purpose in life is to be able to express $h_{0i}$ in terms of $h_{ij}^T$  and $\l$ which are the degrees of freedom of the theory.  For that, we need to get rid of the time derivatives acting in $h_{0i}$. To do so, we take the spatial part of the eom of the scalar field \eqref{scalar_eom}
\bea
&\pd^i h_{ik} + \pd^0 h_{0k} -\dfrac 1 4 \pd_k h  = 0.
\eea
If we differentiate again with respect to the time component we get
\be
\pd_0 \pd^i h_{ik} + \pd_0 \pd^0 h_{0k} -\dfrac 1 4 \pd_0 \pd_k h =0 \label{spatial B},
\ee
so substituting \eqref{spatial B} in the equation for $h_{0i}$ we finally obtain
\bea
& \partial_j \partial^j h_{0i} + m^2 h_{0i} = \pd_0 \pd^j h_{ji}^T.
\eea
With these equations, we can express $h_{0i}$ in terms of the degrees of freedom of the theory.
\par
Finally, let us summarize and compare the results we obtain with the ones in \cite{Chamseddine2,Chamseddine3}. We are studying {\em WTDiff} theories, where the trace of the eom can be gauge fixed using the Weyl symmetry (which is a direct consequence of the change $\hat{g}_{\m\n} = g^{-1/4} g_{\m\n}$). Therefore, after redefining the metric, all the construction possesses this symmetry and nothing breaks it. Moreover, we know that Unimodular Gravity ({\em TDiff}) and General Relativity are equivalent theories at the level of the equations of motion when Bianchi identities are used to reintroduce the trace into the unimodular eom (see Appendix \ref{A}). Up to this point, we expected the same results in the counting of the degrees of freedom in \cite{Chamseddine2,Chamseddine3}. However, the main difference between {\em WTDiff} and {\em Diff} invariant theories lies in the character of the trace of the eom. Even if the form of the eom for both theories is the same after using the Bianchi identity (and when fixing the Weyl gauge, which we can always do), in {\em WTDiff} the trace is related to the integration constant appearing when we integrate the Bianchi identity. As it is seen in our computations, the mimetic degree of freedom, which is related to the trace of the eom, is not dynamical when the symmetry of the theory is {\em WTDiff}. Instead of an oscillating solution for the mimetic degree of freedom in \cite{Chamseddine2, Chamseddine3}, the integration constant is forced to be zero yielding the vanishing of the mimetic degree of freedom on-shell. 
%%%%%%%%%%%%%%%%%%%%%%%%%%%%%%%%%%%%%%%%%%%%%%%%%%%%%%

%%%%%%%%%%%%%%%%%%%%%%%%%%%%%%%%%%%%%%%%%%%%%%%%%%%%%%%%%%%%%%%%%%%%%%%%%%%%%%
\section{Conclusions}
%%%%%%%%%%%%%%%%%%%%%%%%%%%%%%%%%%%%%%%%%%%%%%%%%%%%%%%%%%%%%%%%%%%%%%%%%%%%%%%%    
In the standard reference \cite{Dubovsky} a systematic study is made of the most general possible mass term in the linear approximation that preserves three-dimensional euclidean invariance, namely
\be
L_m\equiv{M_p^2\over 2}\left(m_0^2\, h_{00}^2+ 2 m_1^2\,\sum_i h_{0i}^2-m_2^2\,\sum_{ij} h_{ij}^2+ m_3^2\,\left(\sum_i h_{ii}\right)^2-2 m_4^2\, h_{00}\,\sum_i h_{ii}\right).
\ee    
Before gauge fixing, the massive gravity studied in the present paper corresponds to the mass term
\bea 
L_m\equiv {m^2\over 8}\left(\frac{h^2}{2}-h^{\m\n}h_{\m\n}\right)={m^2\over 8}\left(2h_{0i}h_{0i}+\frac{1}{2}h_{ii}h_{jj}-h_{ij}h_{ij}\right),
\eea
where the constraint $h_{00}=0$ is already implemented. Note that this $h_{\m\n}$ is not traceless yet. This corresponds in the language of \cite{Dubovsky} to the phase
\bea
&m_0^2=0\nonumber\\
&m_1^2=m_2^2=\frac{m_3^2}{2}> 0\nonumber\\
&m_4^2=0.
\eea
This means that this mass term corresponds to the second subcase of (3.25) in \cite{Dubovsky}, which has been argued to be free of pathologies, that is, 
\bea
&m_0^2=0\nonumber\\
&m_1^2, m_2^2> 0\nonumber\\
&(m_1^2-m_4^2)m^2_4=0\nonumber\\
&m^2_2\neq m^2_3.
\eea
\par
To summarize the content of the model, in the {\em WTDiff} theory we are left with the five massive degrees of freedom for the graviton, $h_{ij}^T$, and with a non-dynamical mimetic degree of freedom $\l$ that vanishes on-shell. 
%The amount of mimetic dark matter will depend on initial conditions used to integrate the Bianchi identity that reintroduces the trace into the gravitational eom. 
In \cite{Chamseddine2}, $\l$ is a dynamical degree of freedom with an oscillatory solution which is then argued to be a candidate for cold dark matter. The difference arising in our work is that the dark-matter-like degree of freedom is lost on-shell. Let us stress that this is the first case (that we are aware of) where UG and GR are not fully equivalent at the classical level because of the counting of the dynamical degrees of freedom. The reason for this discrepancy is the relation of the mimetic degree of freedom with the trace of the gravitational eom. That is precisely the critical difference between GR and UG: in UG the trace is not dynamical but a fixed integration constant coming from the integration of the Bianchi identity. Moreover, in this particular model, the constant is fixed to zero so that the mimetic degree of freedom vanishes on-shell.
\par
On the phenomenological side, there is a key difference between the cosmological implications of this model and the standard mimetic massive gravity \cite{Solomon:2019qgf}, namely the loss of the dark-matter-like degree of freedom.
\par
In conclusion, provided one is willing to give up Lorentz invariance, we can get a massive graviton in UG. 
Even if Lorentz invariance is lost, it is interesting that a ghost-free massive deformation of unimodular gravity does exist at all.

%%%%%%%%%%%%%%%%%%%%%%%%%%%%%%%%%%%%%%%%%%%%%%%%%%%%%%%%%%%%%%%%%%%%%%%%%%%%%%
\section*{Acknowledgments}
%%%%%%%%%%%%%%%%%%%%%%%%%%%%%%%%%%%%%%%%%%%%%%%%%%%%%%%%%%%%%%%%%%%%%%%%%%%%%%%%    
GMB is supported by the project FPA2015-65480-P and RSG is supported by the Spanish FPU Grant No. FPU16/01595. This work has received funding from the European Unions Horizon 2020 research and innovation programme under the Marie Sklodowska-Curie grants agreement No 674896 and No 690575. We also have been partially supported by FPA2016-78645-P(Spain), COST actions MP1405 (Quantum Structure of Spacetime) and  COST MP1210 (The string theory Universe). This work is supported by the Spanish Research Agency (Agencia Estatal de Investigacion) through the grant IFT Centro de Excelencia Severo Ochoa SEV-2016-0597.

\newpage
%%%%%%%%%%%%%%%%%%%%%%%%%%%%%%%%%%%%%%%%%%%%%%%%%%%%%%%%%%%%%%%%%%%%%%%%%%%%%%%%%%%%%%%%%%%%%
\appendix
\section{Equivalence of the classical equations of motion} \label{A}
The standard argument that  UG ({\em TDiff}) is equivalent to GR ({\em Diff}) (except for the character of the cosmological constant that appears as a constant of integration in UG) only holds for the unimodular metric, $\hat{g}_{\a\b}$.
The eom that follow from the action
\be
S=\int d^n x \hat{R}
\ee
are
\be
\hat{R}_{\m\n}-{1\over n} \hat{R} \hat{g}_{\m\n}=0.
\label{eomug}
\ee
The contracted Bianchi identities for the unimodular metric
\be
\hat{\nabla}_\m \hat{R}^{\m\n}={1\over 2} \nabla^\n \hat{R}
\ee
implies that
\be
\hat{R}_{\m\n}-{1\over 2}\left(\hat{R}+ 2 \l\right)\hat{g}_{\m\n}=0,
\label{eomugL}
\ee
where
\be
\l\equiv -{n-2\over 2n}\hat{R}.
\ee
As has been pointed out in \cite{AlvarezGMM}, when using an unrestricted via the Weyl transformation $\hat{g}_{\m\n}= g^{-1/n} g_{\m\n}$ the UG action principle reads
\be
S_{\text{\tiny WTDiff}}\equiv -M_p^{n-2}\int d^n x\, |g|^{1/n}\bigg\{R+{(n-1)(n-2)\over 4 n^2 }\, \dfrac{\N_\m g \N_\n g}{g^2}  g^{\m\n}\bigg\}
\ee
and the corresponding equations of motion read
\bea
R_{\m\n}-{1\over n} R g_{\m\n}&={(n-2)(2n-1)\over 4 n^2}\left({\N_\m g\N_\n g\over g^2}-{1\over n}{\N_\m g \N^\m g\over g^2}g_{\m\n}\right)\nonumber\\
&-{n-2\over 2n}\left(\dfrac{\N_\m\N_\n g}{g}-\dfrac{1}{n} \dfrac{\Box g}{g}g_{\m\n}\right),
\label{eomg}
\eea
where both the action as well as the equations  are indeed Weyl invariant ({\em WTDiff}). 
It can be checked that these equations of motion can be obtained via a Weyl transformation of the unimodular eom \eqref{eomug}. After using the Bianchi identity in the unimodular eom \eqref{eomug}, we obtain Einstein's equations with a cosmological constant for unimodular metrics \eqref{eomugL}

%We can Weyl transform that equation so that we get
\bea
&\hat{R}_{\mu \nu} - \dfrac{1}{2} (\hat{R} -2 \lambda) \hat{g}_{\mu\nu} = 0 \quad \rightarrow \nonumber \\
&R_{\mu\nu} \ - \ \dfrac{1}{2} R g_{\m\n} \ - \  \l g^{-1/n} g_{\mu \nu} \ - \ \dfrac{(n-2)(2n-1)}{4 n^2} \dfrac{\N_\m g \N_\n g}{g^2} \ + \dfrac{(n-2)(5n-3)}{8n^2}  \dfrac{\N_\a g \N^\a g}{g^2} g_{\m\n} \ \nonumber \\
&+ \ \dfrac{(n-2)}{2n}  \dfrac{\N_\m \N_\n g}{g} \ - \ \dfrac{(n-2)}{2n} \dfrac{\Box g}{g} g_{\m\n} = 0,
\label{eomwL}
\eea
where $\l$ takes the form 
\be
\l = - \dfrac{(n-2)}{2n} \hat{R} =-g^{1/n} \left( \dfrac{(n-2)}{2n} R- \dfrac{(n-2)(n-1)(5n-2)}{8n^3} \dfrac{ \N_\m g \N^\m g}{g^2}  + \dfrac{(n-1)(n-2)}{2n^2}  \dfrac{\Box g}{g}   \right) .
\ee
Of course, we can always gauge fix $g=1$ to recover the UG results. 

\end{document}